\documentclass[a4paper,fleqn]{cas-sc}

\usepackage[square, numbers, sort&compress]{natbib}
\usepackage{amsmath}
\usepackage{slashed}

\newcommand{\Lagrangian[1]}{\mathcal{L}_{#1}}

\def\tsc#1{\csdef{#1}{\textsc{\lowercase{#1}}\xspace}}
\tsc{WGM}
\tsc{QE}

\begin{document}
\let\WriteBookmarks\relax
\def\floatpagepagefraction{1}
\def\textpagefraction{.001}

\shorttitle{Optimal search reach for heavy neutral leptons at a muon collider}    

\shortauthors{K. M\k{e}ka{\l}a, J. Reuter, A.F. \.Zarnecki}  

\title [mode = title]{Optimal search reach for heavy neutral leptons at a muon collider}  

\author[1,2]{Krzysztof M\k{e}ka{\l}a}[orcid=0000-0003-4268-508X]
\cormark[1]
\ead{k.mekala@uw.edu.pl}
\affiliation[1]{organization={Deutsches Elektronen-Synchrotron DESY},
            addressline={Notkestr. 85}, 
            city={Hamburg},
         citysep={},
            postcode={22607}, 
            country={Germany}}

\affiliation[2]{organization={Faculty of Physics, University of Warsaw},
            addressline={Pasteura 5}, 
            city={Warszawa},
         citysep={},
            postcode={02-093}, 
            country={Poland}}
            
\author[1]{J\"urgen Reuter}[orcid=0000-0003-1866-0157]
\ead{juergen.reuter@desy.de}
\author[2]{Aleksander Filip \.Zarnecki}[orcid=0000-0001-8975-9483]
\ead{filip.zarnecki@fuw.edu.pl}

\cortext[1]{Corresponding author}

\begin{abstract}
 Neutrinos are the most elusive particles known. Heavier sterile neutrinos mixing with the standard neutrinos might solve
 the mystery of the baryon asymmetry of the universe. In this letter, we show that among all future energy frontier accelerators, muon colliders will provide the farthest search reach for such neutrinos for mass ranges above the $Z$ pole into the multi-TeV regime, becoming the optimal machine for this kind of studies. We compare the performance of muon with electron colliders of the same machine energy and briefly discuss the complementarity in flavor space between the two types of accelerators. 
\end{abstract}




\maketitle

\paragraph{Introduction}
\label{sec:intro}

Massive neutrinos are considered the first established building blocks of physics beyond the Standard Model (SM) of particle physics. In several extensions of the fundamental theory, their tiny masses are attributed to originate from seesaw-like mixing with heavier sterile neutrinos whose masses could be all the
way from the electroweak (EW) to the unification scale. While long-distance neutrino oscillation experiments like DUNE or Hyper-Kamiokande will shed more
light on the mass hierarchy and the mixing parameters, heavier neutrinos can be directly searched for at hadron colliders such as the
LHC and future lepton colliders \cite{Petcov:1984nf, Dittmar:1989yg, Sirunyan:2018mtv, ATLAS:2019kpx, LHCb:2020wxx, CMS:2022fut, Pascoli:2018heg, delAguila:2005pin, delAguila:2005ssc, Saito:2010xj, Das:2012ze, Banerjee:2015gca, Antusch:2016ejd, Cai:2017mow, Chakraborty:2018khw, Das:2018usr, Ding:2019tqq,  Mekala:2022cmm, Lu:2022wsm, Bose:2022obr, Huber:2022lpm}. For the simplest extension of the Standard Model comprising three heavy neutral leptons (of which we consider only the lightest), three different regimes can be probed at colliders:
light neutrinos which are long-lived and result in displaced vertices or decay outside the detectors, intermediate-mass neutrinos that
decay promptly and are dominantly produced in $Z$ (and $W$ or Higgs) decays, and heavy neutrinos with masses $m_N$ $ \gtrsim M_H$. In this paper,
building upon an analysis framework similar to earlier studies for searches at linear $e^+e^-$ machines~\cite{Mekala:2022cmm}, we focus on the third case and show that the most sensitive searches for direct single) heavy neutrino production are possible at high-energy muon colliders.\footnote{We do not consider models with extended gauged groups in which pair production of heavy neutrinos or neutrinos from heavy gauge boson decays might dominate.} Lepton colliders are, in general, sensitive to much
smaller mixing parameters and hence to much higher scales of UV completions. We will consider a muon collider setup
with energies of 3 and 10 TeV, and integrated luminosities of 1 and 10 ab${}^{-1}$, respectively \cite{Delahaye:2019omf, Long:2020wfp, Black:2022cth}.

\paragraph{Model setup and simulation framework}
In this letter, we consider the Phenomenological Type I
Seesaw Mechanism~\cite{delAguila:2008cj, Atre:2009rg}, implemented within the \textit{HeavyN} model with
Dirac neutrinos~\cite{HeavyN, Degrande:2016aje, Pascoli:2018heg}, {\em i.e.} we assume it just as a representative model candidate without any prejudice (our findings are quite generic, though specific model setups like artificial flavor mixings could of course lead to singular cases; Refs.~\cite{Kilian:2003xt,Schmaltz:2004de} provide an example where such heavy neutrinos appear even at a multi-TeV scale UV completion). This effective extension of the SM introduces three flavors of right-handed neutrinos (denoted as $N_k$) that are singlets under the SM gauge groups. The Lagrangian of the model reads:
\begin{equation}
\Lagrangian[] 
= \Lagrangian[SM] + \Lagrangian[N] + \Lagrangian[WN\ell] + \Lagrangian[ZN\nu] + \Lagrangian[HN\nu]
\end{equation}
where $\Lagrangian[N]$ is a sum of kinetic and mass terms for heavy neutrinos (in 4-spinor notation, which combines terms with spinors of dotted and undotted indices):
\begin{equation}
\Lagrangian[N] = \bar{N}_{k}i\slashed{\partial}N_{k} - m_{N_{k}}\bar{N}_{k}N_{k} \qquad\textnormal{for}\; k = 1,\,2,\,3,
\end{equation}
$\Lagrangian[WN\ell]$ yields neutrino interactions with the $W$ boson:
\begin{equation}
\Lagrangian[WN\ell] = - \frac{g}{\sqrt{2}}W^{+}_{\mu} \sum^{3}_{k=1}\sum^{\tau}_{l=e}\bar{N}_{k}V^{*}_{lk}\gamma^{\mu}P_{L}\ell^{-} + \textnormal{ h.c.},
\end{equation}
$\Lagrangian[ZN\nu]$ interactions with the $Z$ boson:
\begin{equation}
\Lagrangian[ZN\nu] = - \frac{g}{2\cos\theta_{W}}Z_{\mu} \sum^{3}_{k=1}\sum^{\tau}_{l=e}\bar{N}_{k}V^{*}_{lk}\gamma^{\mu}P_{L}\nu_{l} + \textnormal{ h.c.},
\end{equation}
and $\Lagrangian[HN\nu]$ interactions with the Higgs boson:
\begin{equation}
\Lagrangian[HN\nu] = - \frac{gm_{N}}{2M_{W}}h \sum^{3}_{k=1}\sum^{\tau}_{l=e}\bar{N}_{k}V^{*}_{lk}P_{L}\nu_{l} + \textnormal{ h.c.}
\end{equation}
The UFO library of the model contains 12 free parameters in addition to the SM parameters, which are three masses of the heavy neutrinos m$_{N_k}$ and nine real (no $CP$ violation assumed for simplicity) mixing parameters $V_{lk}$, where $l = e, \mu, \tau$ and $k = N_1, N_2, N_3$. For the purpose of this analysis, we considered a scenario with only one heavy Dirac neutrino $N_1 \equiv N$ with a mass below $\mathcal{O}(30\,\text{TeV})$ and equal couplings to all SM leptons ($|V_{eN_1}|^{2} = |V_{\mu N_1}|^{2} = |V_{\tau N_1}|^{2} \equiv V_{l N}^{2}$). For reference sample generation, the mixing parameter $V_{lN}^{2}$ has been set to 0.0003. Other values for the mixing parameters in the analysis below were accessed via rescaling with the corresponding cross section. Although there are many different possible signatures of such particles at future colliders, for center-of-mass energies above the $Z$ pole, the t-channel $W$ exchange resulting in the production of a light-heavy neutrino pair ($\mu^+ \mu^- \rightarrow N \, \nu$) is one of the most promising production channels~\cite{Antusch:2016ejd} and the production cross section is of the order of 1-10 fb for masses of the neutrinos up to the collision energy. For our choice of the parameter space, the heavy neutrino has a microscopic lifetime ($c\tau \ll 1$ nm) and no displaced vertices are expected. Among the possible decay channels of these particles, only the signature of two jets and a lepton ($N \rightarrow qq\ell$) allows for direct reconstruction of the mass of the heavy state. However, one can also consider other production and decay channels, for example, those induced by vector-boson fusion of a neutrino pair or resulting in the mono-Higgs production and decay which could potentially offer large cross sections at a high-energy lepton machine (see e.g. \cite{Costantini:2020stv}). We decided to limit our study to the $qq\ell \nu$ signature only, which gives direct access to the heavy-neutrino kinematics, and prove by itself the superiority of the Muon Collider over accelerators of other kinds in searches for very massive neutrinos. The design of the more general search procedure, involving the interplay between different production and decay channels and requiring the development of a proper analysis procedure for each of them, and finally their combination, has been left for further studies.

\begin{figure}
    \centering
\includegraphics[width=0.7\textwidth]{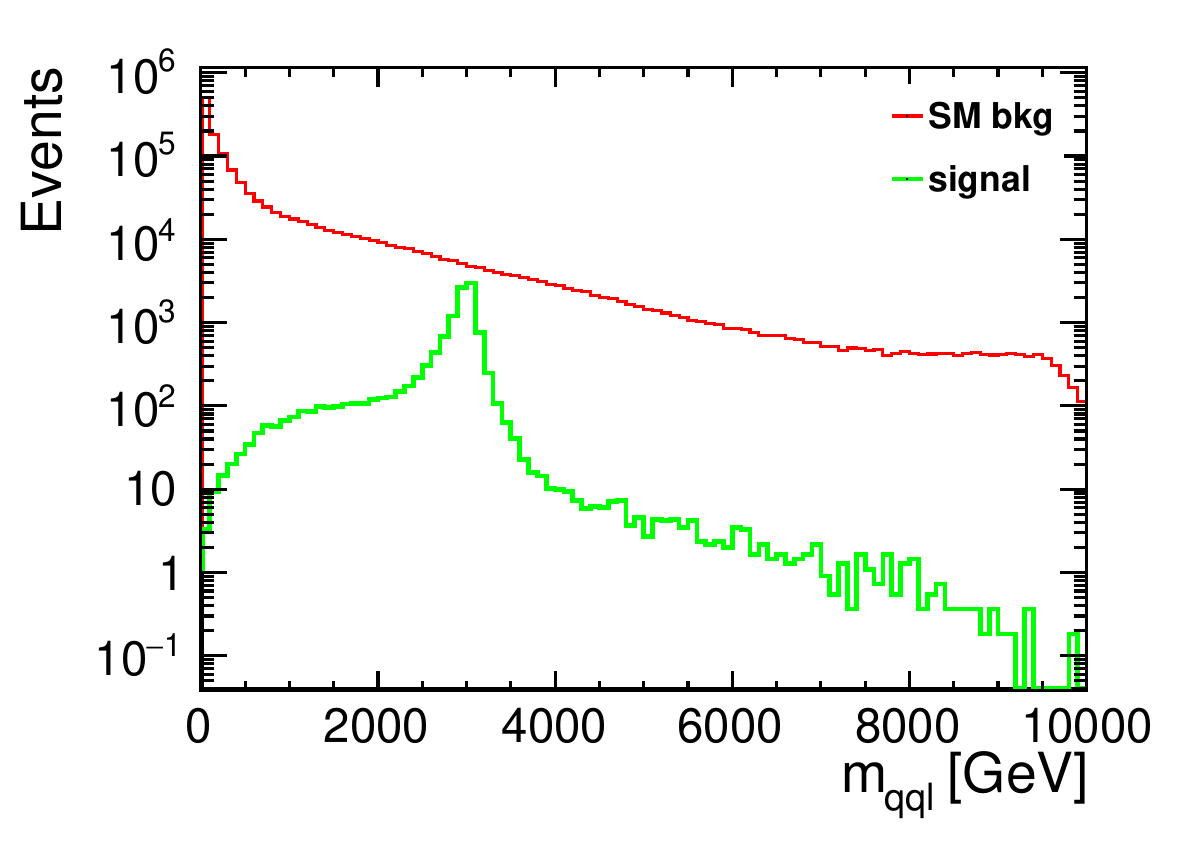}
    \caption{$qq\ell$ mass distribution for a reference scenario assuming the existence of one Dirac neutrino with a mass of 3\,TeV, at a 10\,TeV muon collider. The red line stands for the $\mu^+\mu^-$ background and the thick green one for the signal scenario.}
    \label{fig:mass}
\end{figure}

In the first step, we generated event samples with \textsc{Whizard 3.0.2}~\cite{Moretti:2001zz,Kilian:2007gr,Brass:2018xbv} at leading order (LO) in the SM coupling constants (although recently higher-order corrections have become available in an automated manner~\cite{Bredt:2022dmm}), while parton showering and hadronization were performed using the built-in interface to \textsc{Pythia~6}~\cite{Sjostrand:2006za}. Then, we simulated detector response with \textsc{Delphes 3.5.0}~\cite{deFavereau:2013fsa} using default Muon Collider detector cards.
At the generator level, a set of cuts was applied to remove possible singularities. They included 10-GeV cuts on the energy of produced jets and leptons, the invariant mass of quark and lepton pairs, and the four-momentum transfer from  the incoming muons. Furthermore, it was required that at least one lepton could be detected in the central detector (we assumed $5^{\circ}<\theta<175^{\circ}$, where $\theta$ is the lepton polar angle).
For the detector simulation, the VLC clustering algorithm in the exclusive two-jet mode (R = 1.5, $\beta$ = 1, $\gamma$ = 1 -- see \cite{Boronat:2016tgd}) was applied. The choice of the mode corresponds to the expected signal topology, consisting of two reconstructed jets and one lepton.
Since the considered \textsc{Delphes} model cannot generate fake lepton tracks, only 4- and 6-fermion background processes with at least one lepton in the final state ($qq \ell \nu$, $qq \ell \ell$, $\ell \ell \ell \ell$, $qqqq \ell \nu$, $qqqq \ell \ell$, $qq \ell \nu \ell \nu$, $qq \ell \nu \nu \nu$) were generated. The most important channels in terms of cross section ($\mathcal{O}(1\textnormal{ ab})$ at both energy stages) were $qq \ell \nu$ and $\ell \ell \ell \ell$; the latter could be, however, easily reduced by lepton identification. Background channels induced by photons from collinear initial-state splittings 
were neglected, as their impact on the final results was found to be marginal.

\paragraph{Analysis procedure}
In the next step, a set of selection cuts was applied to reject events incompatible with the expected topology of two jets and one lepton.
To exclude events with significant contributions of forward deposits assigned to the beam jets, an upper limit of 20 GeV was applied on the transverse momentum of objects not included in the reconstructed final state. In Figure \ref{fig:mass}, we show a distribution of the invariant mass of two jets and a lepton for a reference scenario (a 3\,TeV neutrino at a 10\,TeV muon collider). A peak corresponding to the mass of the heavy neutrino is clearly visible. The left tail is due to events with $\tau$ decays, for which the escaping neutrinos reduce the reconstructed invariant mass. On the right-hand side, the tail is an effect of finite detector resolution.
\begin{figure}
    \centering
    \includegraphics[width=0.7\textwidth]{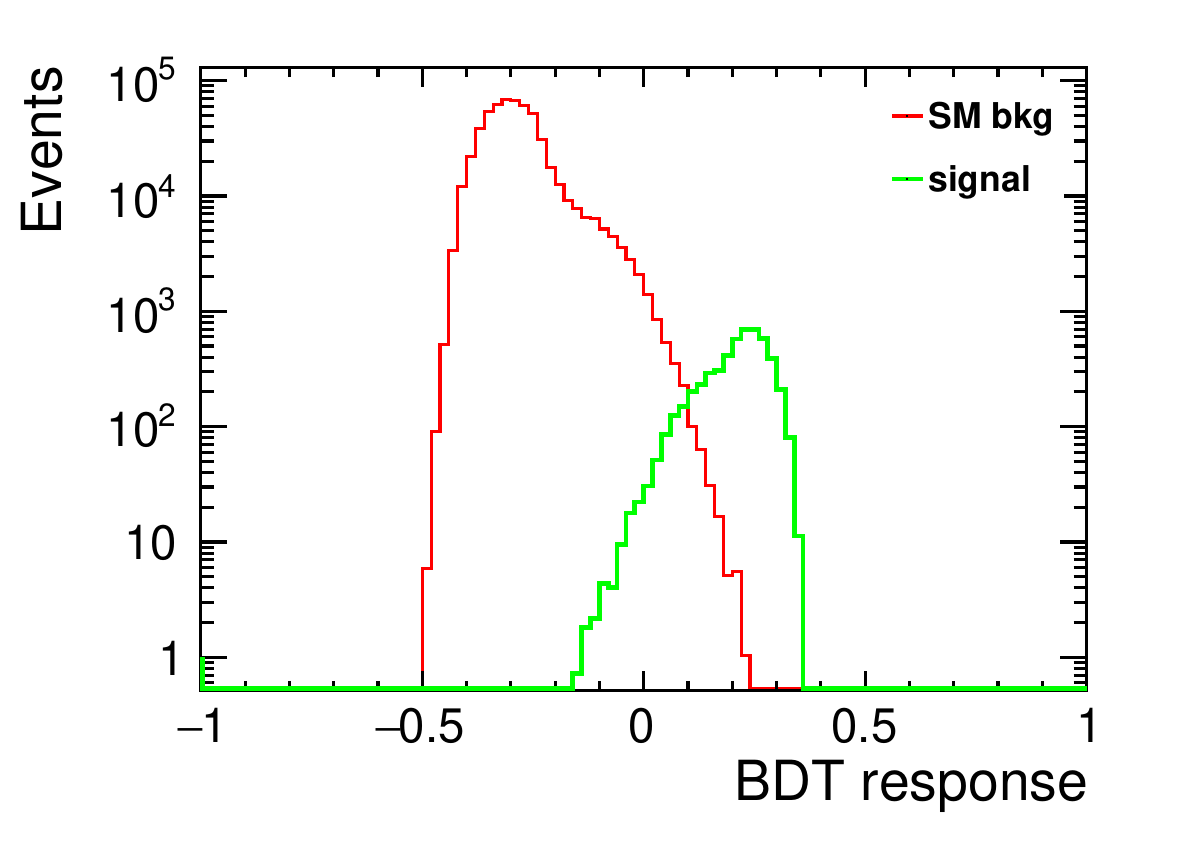}
    \caption{Distribution of the BDT response for the reference scenario (Dirac neutrino, m$_N$ = 3\,TeV) with electrons in the final state at a 10\,TeV muon collider. The red line denotes the background, and the green line the signal.}
    \label{fig:BDT_response}
\end{figure}
Subsequently, we applied the Boosted Decision Tree (BDT) method implemented in the TMVA package~\cite{Hocker:2007ht} to discriminate between signal and background events. A set of eight variables describing event kinematics was chosen to optimize the classification:
\begin{itemize}\setlength\itemsep{-0.2em}
\item m$_{qq\ell}$ -- invariant mass of the dijet-lepton system,
\item E$_{qq\ell}$ -- energy of the dijet-lepton system,
\item p$^{T}_{qq\ell}$ -- transverse momentum of the dijet-lepton system,
\item $\alpha$ -- angle between the dijet system and the lepton,
\item $\alpha_{qq}$ -- angle between the two jets,
\item p$^{T}_{qq}$ -- dijet transverse momentum,
\item E$_{\ell}$ -- lepton energy,
\item p$^{T}_{\ell}$ -- lepton transverse momentum.
\end{itemize}
In the choice of the BDT variables, we followed the approach of Ref. \cite{Mekala:2022cmm}. In general, variables involving two jets and a lepton allow for direct reconstruction of the heavy state. On the other hand, variables describing the di-jet and the lepton separately serve to differentiate between background and signal kinematics. Their combination within the BDT procedure helps determine if the reconstructed particles come from the decay of a new heavy state or of a known SM particle. It was found that a larger set of variables would not significantly improve the presented results but would increase the computation time. Due to the considerable difference between the composition of the expected background for the two channels, the algorithm was implemented separately for events with reconstructed electrons and muons in the final state. The BDT response for the reference scenario is shown in Figure \ref{fig:BDT_response}. The two distributions confirm that a very efficient separation of signal and background events is possible. The distributions were used to extract the expected limits on the coupling parameter $V^2_{lN}$ within the CL$_s$ method, implemented in the \textit{RooStats} package~\cite{Moneta:2010pm}. This allowed for combining the electron and muon channels. The impact of systematic uncertainties has been neglected at this stage, as they are not expected to affect the final conclusions significantly.

\begin{figure}
    \centering
    \includegraphics[width=0.7\textwidth]{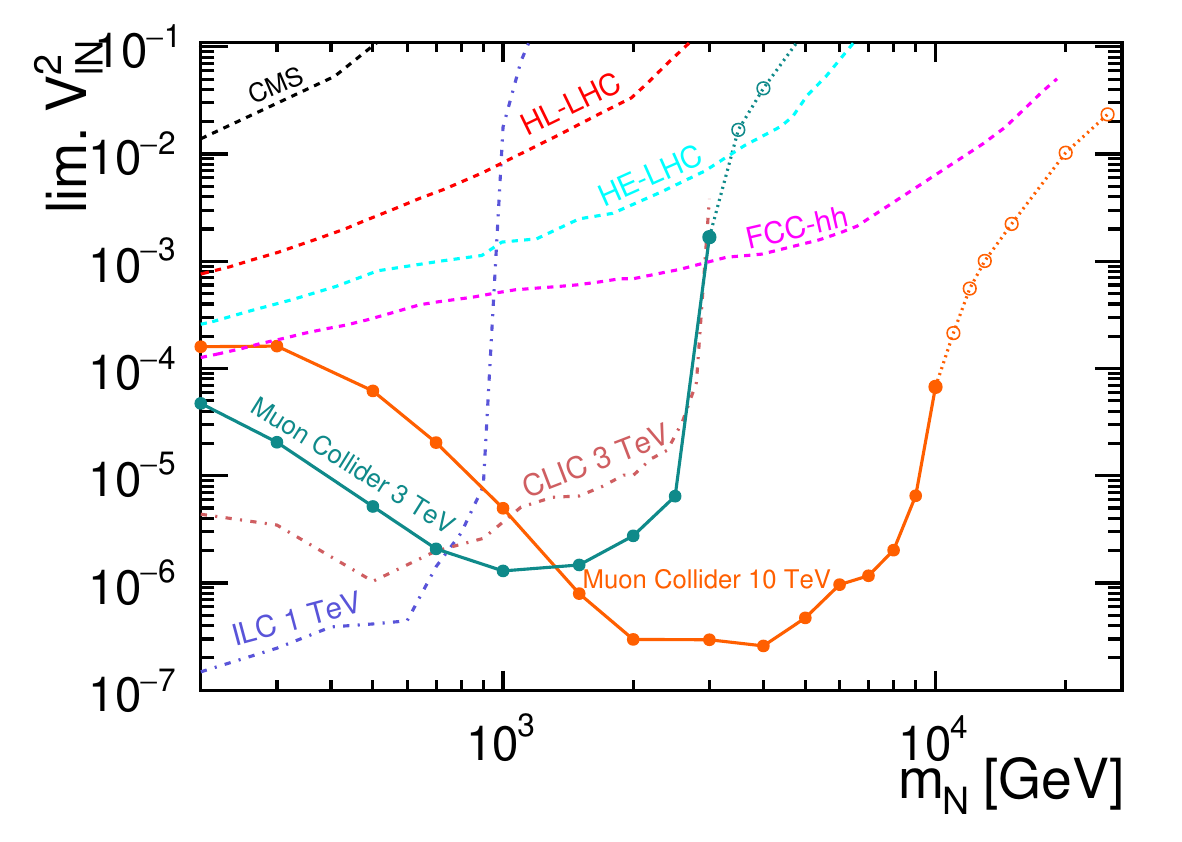}
    \caption{Limits on the coupling $V^2_{\ell N}$ for different Muon Collider setups (3 TeV, 1 ab$^{-1}$ -- turquoise; 10 TeV, 10 ab$^{-1}$ -- orange) resulting from the search for single on-shell (solid line) and off-shell (dotted line) heavy neutrino production. Dashed lines indicate limits~\cite{Sirunyan:2018mtv, Pascoli:2018heg, Mekala:2022cmm} from current and future hadron machines (current CMS limits, 13 TeV, 35.9 fb -- black; HL-LHC 14 TeV, 3 ab$^{-1}$ -- red; HE-LHC 27 TeV, 15 ab$^{-1}$ -- cyan; FCC-hh 100 TeV, 30 ab$^{-1}$ -- pink), dashed-dotted for $e^+e^-$ colliders (ILC 1 TeV, 3.2 ab$^{-1}$ -- violet; CLIC 3 TeV, 4 ab$^{-1}$ -- coral).}
    \label{fig:results}
\end{figure}

\paragraph{Results}
In Figure \ref{fig:results}, limits on the coupling $V^2_{lN}$ for the two Muon Collider setups (3 TeV, 1 ab$^{-1}$ and 10 TeV, 10 ab$^{-1}$) are presented and compared with the current limits coming from the CMS experiment (Majorana neutrinos, Fig. 2 in \cite{Sirunyan:2018mtv}), as well as with the expectations for future hadron colliders (Dirac neutrinos, Fig. 25b in \cite{Pascoli:2018heg}) and $e^+e^-$ colliders (Dirac neutrinos, Fig. 12 in \cite{Mekala:2022cmm}). It should be noted that in the hadron collider analyses, heavy neutrino decays into taus were not considered, and thus their sensitivity is enhanced relative to the results presented for the lepton colliders, where the tau-channel decays are included. 
Also included in Figure \ref{fig:results} are results for the off-shell heavy neutrino production (indicated with dotted lines). In this case, simple quadratic scaling of the reference-scenario cross section to other coupling values is no longer valid and one should numerically search for the proper coupling value. Nevertheless, for relatively narrow resonances (as the one considered in this letter), the cross section grows faster with the neutrino width (which is always quadratically related to the mixing value) than in the on-shell case and thus, the results in this region obtained within our framework can be considered as conservative. It is remarkable then that the presented results surpass those for the hadron machines even for heavy-neutrino masses above the collision energy when the new particles could be produced only off-shell.
As shown in Figure \ref{fig:results}, limits expected from the $e^+e^-$ colliders, ILC running at 1\,TeV and CLIC running at 3\,TeV, are more stringent for masses of the heavy neutrinos up to about 700\,GeV. The fact that the results for CLIC and a Muon Collider operating at the same energy of 3 TeV do not coincide may be surprising. However, several effects must be taken into account for a proper comparison: the most important factors are different integrated luminosities and beam polarizations. In addition, the beam spectra and the beam-induced background channels cannot be neglected for $e^+e^-$ colliders, while their impact is significantly reduced for $\mu^+\mu^-$ machines. It was verified that, for the same generation setup (no beam polarization, no beam spectrum, no beam-induced background channels, but different initial-state particles and detector designs), the expected CLIC limits are consistent with the Muon Collider ones, giving the analysis precision. The discrepancy visible in Figure \ref{fig:results} could then be explained as follows: at lower neutrino masses, the expected limits from CLIC are more stringent due to the higher integrated luminosity and electron beam polarization, and at higher masses, they are worse because of the impact of the luminosity spectra and beam-induced backgrounds.

In the analysis, we assumed that all the mixing parameters $V_{lN}$ have the same value. It is important to note that this approach is not unique. Using data from both electron-positron and muon colliders, one could potentially loosen this assumption and constrain the parameters $V_{eN}$ and $V_{\mu N}$ separately, by either excluding tau coupling from the physical model or implementing a proper tau tagging procedure to constrain it. Such a method would give limits not only on the couplings themselves but also on their products in the framework where couplings are treated independently, possibly hinting at a flavor-universality violation. The details are, however, beyond the scope of this letter.

\paragraph{Conclusions} 
Extensions of the Standard Model introducing heavy neutrinos offer interesting solutions to several of its open questions, e.g. the baryon asymmetry of the universe, dark matter and flavor. If such particles are at mass scales well above a GeV, they can be efficiently searched for at future lepton colliders. Due to the highest achievable energies and the clean experimental environments, muon colliders would provide the furthest discovery reach for TeV-scale neutrinos in such kind of models, vastly surpassing high-energy hadron colliders, potentially even for neutrino masses above the available collision energy. By employing the synergy of both different types of lepton machines, electron-positron and muon colliders, different paths in the flavor parameter space of the models could be pursued.
\label{sec:conclusions}

\subsection*{Acknowledgments}
The work was partially supported by the National Science Centre
(Poland) under the OPUS research project no. 2021/43/B/ST2/01778.
KM and JRR acknowledge the support by the Deutsche
Forschungsgemeinschaft (DFG, German Research Association) under
Germany's Excellence Strategy-EXC 2121 "Quantum Universe"-3908333.  This work
has also been funded by the Deutsche Forschungsgemeinschaft (DFG,
German Research Foundation) -- 491245950.

\bibliographystyle{unsrtnat}

\bibliography{bibliography}

\end{document}